# Bit-level Parallelization of 3DES Encryption on GPU

Kaan Furkan Altınok, Afşin Peker
Electrical and Electronics Engineering
Middle East Technical University
{furkan.altinok, afsin.peker}@metu.edu.tr

Alptekin Temizel
Graduate School of Informatics
Middle East Technical University
atemizel@metu.edu.tr

*Abstract*— **Triple DES (3DES) is a standard fundamental encryption algorithm, used in several electronic payment applications and web browsers. In this paper, we propose a parallel implementation of 3DES on GPU. Since 3DES encrypts data with 64-bit blocks, our approach considers each 64-bit block a kernel block and assign a separate thread to process each bit. Algorithm's permutation operations, XOR operations, and S-box operations are done in parallel within these kernel blocks. The implementation benefits from use of constant and shared memory types to optimize memory access. The results show an average 10.70x speed-up against the baseline multi-threaded CPU implementation. The implementation is publicly available at: https://github.com/kaanfurkan35/3DES_GPU**

## I. Introduction

ENCRYPTION plays a vital role in today's digital life. As digital transactions and communications become more prevalent, sensitive information is at greater risk of being compromised. Encryption algorithms are used to protect confidential information by encrypting data traveling in an insecure channel. Even though the most important property of an encryption algorithm is generally its toughness, encryption speed is also becoming important since more and more people are using digital services and the amount of data that has to be encrypted and protected is increasing significantly.

There are several ways to speed up the encryption/decryption operations such as using custom hardware. While this increases the speed, it is a costly solution. On the other hand, by exploiting the parallel nature of encryption algorithms, GPUs can also be used to speed up the operations. Since most algorithms work on plaintext blocks, multiple blocks can be encrypted in parallel using GPUs, drastically increasing the size of data that can be encrypted at a unit of time.

In this paper, we propose a parallel GPU implementation of the Triple-DES (3DES) algorithm using CUDA. Different to the existing solutions which parallelize the operations on a block by block basis, the operations are done on a bit-by-bit basis.
In Section II, we describe the prior implementations of 3DES algorithm. Section III provides a detailed description of 3DES followed by its proposed implementation in Section IV. Experimental results are provided in Section V and concluding remarks are in Section VI.

## II. Prior Work

There are a limited number of previous studies on the parallelization of DES/3DES, on CPUs and GPUs. Swierczewski [1] implemented the 3DES algorithm on GPUs using CUDA with considerable improvements over the CPU versions. Çınar et al. [2] implemented the original DES algorithm using CUDA for lightweight systems. They also observe a significant performance increase over the conventional implementation. Fadhil [3] implemented 3DES using CUDA for watermarking images with concealed information and reported that the parallel implementation is about 6x faster. Beletskyy and Burak [4] parallelized the 3DES algorithm using OpenMP on CPUs by analyzing the data dependencies during the operations and parallelizing the loops accordingly. They reported a 1.95x speed-up on a two-CPU machine.

Most of these implementations parallelize the operations on a text-block scale, except [4], in which the implementation is on CPU. In the proposed approach, the parallelization is done at bit-level and it is optimized to run on GPU. Since most of the operations during the encryption/decryption are shuffling and shifting the bits around, these operations can be done for all bits concurrently, improving the performance of the algorithm considerably.

## III. Triple DES

DES is a symmetric-key algorithm for the encryption of digital data. It was designed by IBM in the 1970s and became a worldwide standard until the 1990s. For this algorithm, the block size is 64 bits, key length is 56 bits, and there are 16 rounds using the 16 48-bit subkeys generated from the original 56-bit key. The algorithm starts with an initial permutation of the 64-bit plaintext. Then the data is split into two halves, and for 16 rounds, they undergo certain operations that scramble the data. After 16 rounds, a final permutation is performed, and the ciphertext is obtained. An illustration of the algorithm can be found in Fig. 1. The decryption of the ciphertext is nearly identical; only the round key order must be reversed.

However, due to its short key length, DES is vulnerable to brute force attacks and was subsequently removed from the standards. This led to the 3DES, where the DES encryption-decryption-encryption is applied to the plaintext sequentially with different keys, increasing the effective key length to 168. This proved to be much more secure than DES, since brute-forcing a 168-bit key will take years, and it is still a NIST standard. In the following sections, steps of the 3DES key generation and encryption/decryption algorithms will be presented. The constant arrays mentioned in the following sections can be found in the appendix section.

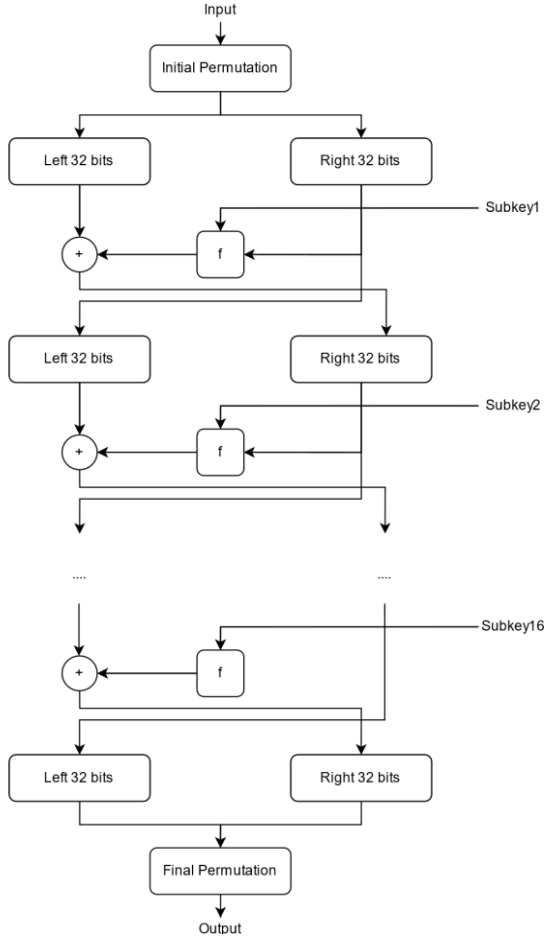

Figure 1: DES algorithm flowchart

### A. Triple-DES Key Generation

Key generation algorithm in 3DES is as follows:
- The 64-bit key is permuted to 56 bits using the *pc_1* array. It has to be noted that 57th bit of the original key will be the 1st bit of the permuted key and so on. The same logic applies for every permutation function (also in encryption).
- The 56-bit key is divided into two 28-bit arrays.
- Each array is circularly left-shifted using the *shift_keys* array in each round (16 total) consecutively.
- A combination of the two arrays, (56 bits), is then permuted to 48-bit to finalize the round key using the *pc_2* array.
- The last two items are repeated 16 times in order to produce 16 round keys. And this whole algorithm is run three times for each base key to achieve 3DES.

### B. Triple DES Encryption/Decryption

Steps of the DES encryption are as follows:
- The 64-bit plain text block is permuted to 64 bits with an *initial_perm* array.
- 64-bit is divided into two arrays of 32 bits each (left and right).
- The round starts here (total 16).
- Right 32 bits are permuted to 48 bits with an *exp_d* array.
- It is XORed with the round key.
- Each 6-bit block from the 48-bit array is supplied to *S-boxes* to reduce it to 4-bits, which finally makes a 32-bit array. Operation is as follows; Six bits are represented as, for example, 010111. Middle 4 bits are 1011, which is 11 in decimal. First and last bits are 01, which makes 2 in decimal. *S-boxes* are accessed with the block number, then row 2 is used, and column 11 is used to access it. The number there will be the 4-bit char array for our algorithm. There are a total of 8 *S-boxes* for each 6-bit block as shown in the appendix.
- The 32-bit array is permuted again to 32 bits with a *per* array.
- Initial left array (which is created before round loops) is XORed with the permuted 32-bit array.
- Left 32-bit and XORed 32-bit are swapped and supplied to the next round.
- Rounds finish here (total 16).
- After 16 loops, 32-bit arrays are combined to form a 64-bit array.
- Final permutation is done from 64-bit to 64-bit using the *final_perm* array.
  - This 64-bit char array is the ciphertext.

Decryption operation also consists of these identical steps, with the only difference that the round key orders are reversed (i.e. the 1st round key for encryption becomes the 16th round key for decryption and so on).

3DES encryption and decryption are implemented as follows, where, $E$ represents encryption operation, $D$ represents decryption operation and the subscripts $Ki$ represent the $i^{th}$ 56-bit key used for these operation:

$$ciphertext = E_{K3}(D_{K2}(E_{K1}(plaintext)))$$
$$plaintext = D_{K1}(E_{K2}(D_{K3}(ciphertext)))$$

This method also allows backward compatibility, since if all three base keys are the same, the system behaves like the original DES.

## IV. THE PROPOSED APPROACH

In this section, we introduce the proposed bit-wise parallelization scheme and explain the memory design for optimized access. Two separate kernels were implemented for key generation and encryption/decryption, which are described in separate sections below.

### A. Key Generation Kernel

Flowchart of the kernel can be found in Fig. 2. This kernel has 3 blocks for each base key, and each block consists of 56 threads. Even though the keys are 64 bits, in the first step, they are permutated to 56 bits and 56 threads, corresponding to each bit are used. The kernel will take an array of 3 base keys (64 bits) as input and will output an array of 48 round keys (48 bits), 16 for each base key. The algorithm will be parallelized as follows:

- Permutation is performed bitwise; each thread reads an element from *pc_1* and uses that value to index the base key bits accordingly. The read bit is written to the permuted 56-bit array for the next steps.
- A thread in each block splits the permuted key into two.
- The final step of the algorithm cannot be fully parallelized as each round has to use the result of the previous round. So, each block runs a for loop for 16 times for each subkey.
- In the for loop, shifts are handled by two threads in each

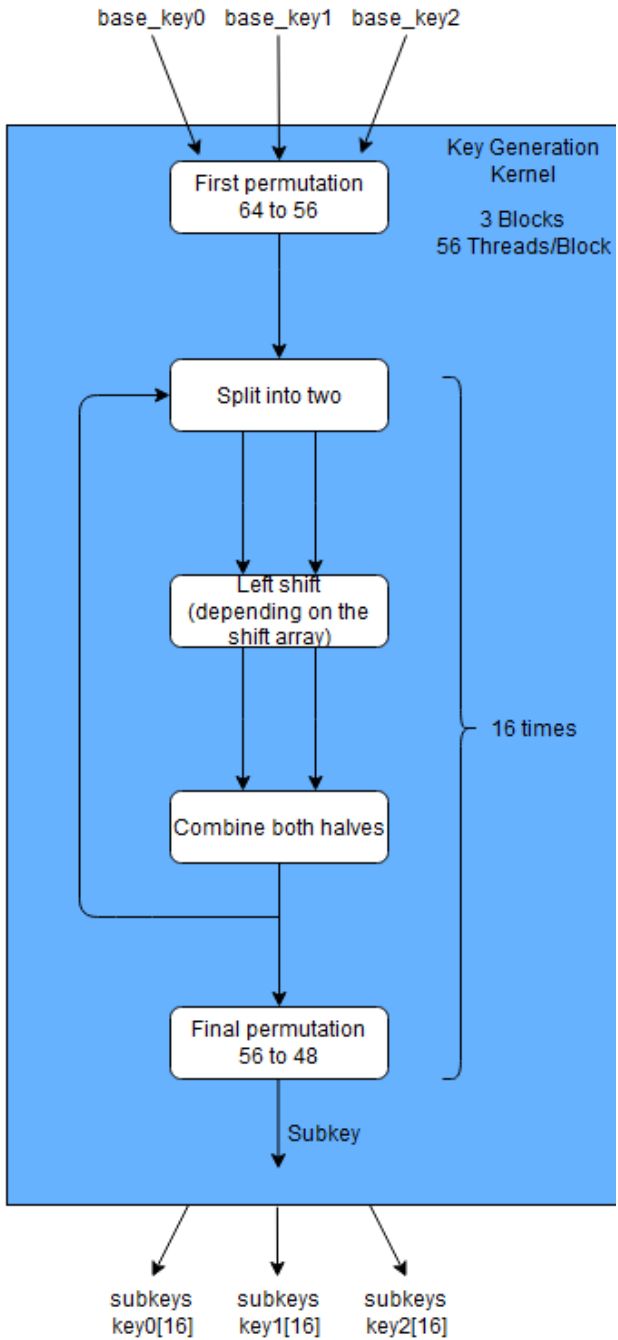

Figure 2: Key generation kernel flowchart

block, for left and right half of the 56-bit permuted key. Then, another permutation is performed using *pc_2*, in the same fashion as in the first step. The output of this permutation is one of the subkeys, it is written to the output, and this operation is repeated 16 times.

*B. Encryption/Decryption Kernel*

After key generation kernel, encryption/decryption kernel is run. Flowchart of the kernel is shown in Fig. 3. This kernel has as many blocks as the plaintext, and each block consists of 64 threads since the blocks in DES are 64 bits. In other words, each kernel block is responsible for a block of plaintext, and each thread is responsible for a bit in the plaintext block. It will take the plaintext, subkeys, and mode as the input. Depending on the mode (encryption or decryption), relevant subkeys are used and the output is the encrypted/decrypted text. Parallelization of the algorithm is as follows:

- The first permutation using the *initial_perm* array is performed in parallel, like in the key generation kernel.
- A thread in each block splits the permuted text into two. This part of the algorithm cannot be parallelized fully, as in the case of key generation. So, the following operations are executed in a loop for 16 rounds. In each iteration of the loop, the first 48 threads in each block expands the right half to 48 bits. The expanded half is XORed with the round key, each bit in processed in parallel.
- Next, 6-bit groups of the expanded and XORed right half use the *S-boxes* in parallel. There are eight such groups, and this operation is handled by four threads for each group. One of these four threads accesses the *S-box* and gets the substitution value, and four threads write each bit of this value to the respective indices of the substituted array.
- This array is permutated in parallel using the *per* array.
- The output of this permutation is XORed with the left half, each bit in parallel.
- Finally, the left half and the output of the previous step are swapped. These operations are repeated 16 times.
- After 16 rounds, two 32-bit arrays are combined to get a 64-bit array.
- The *final_perm* array is sed for a final permutation of this 64-bit array in parallel, and the output is the encrypted/decrypted text block, depending on the mode.

This kernel needs to be called three times consecutively for 3DES. For decryption, the same kernel is used by first reversing the supplied key order on the CPU before the kernel launch.

*C. Memory Design*

A text file is provided to the main function, which includes a plain text in any size consisting of 64-bit blocks, three different 64-bit keys as characters. Since CPU and GPU registers cannot fetch at bit granularity, the smallest available data type (*char*) is used to represent bits. CPU sends the plaintext and keys as *char* arrays to the global memory of the GPU.

Constant tables like permutation tables, expansions tables, S-box tables are stored in the read-only cache. Constant memory is not suitable for these tables since every thread accesses a different index of those tables. On the other hand, shift table is stored in constant memory as every thread accesses its same index at the same time.

Plaintext is divided into 64-bit blocks for each kernel block, and in the kernel, plaintext blocks are copied to the shared memory for processing.

From three different keys, 48 keys are used in each 3DES round. Each thread accesses a different bit of the key, so again, the read-only cache is the preferred memory type to store the keys.

## V. EXPERIMENTS AND RESULTS

To evaluate the performance of the proposed algorithm, we used text files with different sizes and measured the encryption time of our algorithm on two different GPUs and CPUs: GPU-1: NVIDIA GeForce RTX 2060 Super, GPU-2: NVIDIA GeForce GTX 1660 Super Gaming X, CPU-1: AMD Ryzen 5 3600 (6-cores) and CPU-2: Intel i7 9700K (8-cores). In all cases, encryption function was used in electronic codebook (ECB) mode, that is, each block is encrypted independently from each other. The block sizes were increased exponentially, starting from $2^2$ to $2^{17}$.

For GPU measurements, NSight profiler was used. Encryption kernel average running time was multiplied by three since 3DES consists of 3 kernel launches. For CPU measurements, classical CPU implementation was partly parallelized using OpenMP, for a fair comparison. More specifically, the *omp parallel for* pragma was used on the for loop that iterates over each block of plaintext. Hence, several blocks get encrypted in parallel.

In Table 1, the encryption times for different block sizes are shown for all devices. It can be seen that as the block size increases, the processing time also increases linearly for the CPUs while the increase is sub-linear for the GPUs. While GPUs do not offer any advantage for smaller block sizes, the speed-up advantage becomes more pronounced with larger block sizes. This effect is also shown in Fig. 4.

Table II shows the speed-up of each GPU with respect to each CPU for different block sizes. Table III shows average speed-ups. The results show that the proposed GPU implementation is 9.31 to 10.70 times on GPU-1 and 7.01 to 8.03 times faster on GPU-2 for files bigger than 8 KB. The best speed-up is observed for 64 KB files, with up to 20.25 times.These results show that our GPU implementation speeds up the 3DES algorithm considerably, compared to the conventional CPU implementation. The expected speed-up in real-life cases is expected to be higher than these averages when bigger file sizes are used.

Comparing the results with prior works in the literature was not possible since the results were reported on older hardware setups and it was not possible to access their source codes.

An analysis with NSight Profiler shows that compute and memory are well-balanced in the encryption kernel. Considering memory workload analysis, L1 Cache Hit is 98.69%, which indicates that it was effective to put constant arrays to the read-only cache. Maximum bandwidth utilization is reported to be 71.16% which indicates the implementation was effectively using the bandwidth and there was no bottleneck. Theoretical occupancy is 100%, and theoretical SM (warp/cycle) is 32 ideally. The achieved occupancy reported by the profiler, 98.25%, and achieved active warps per SM, 31.44, were close to these theoretical upper limits.

## VI. Conclusion

The results showed that, especially in big data chunks, modern GPUs can provide significant speed-up against multi-threaded CPU implementation for parallelizable cryptography algorithms. While it was not possible to experimentally compare the proposed implementation with those in the literature, we made our source code publicly available to make our results reproducable and enable future comparisons.

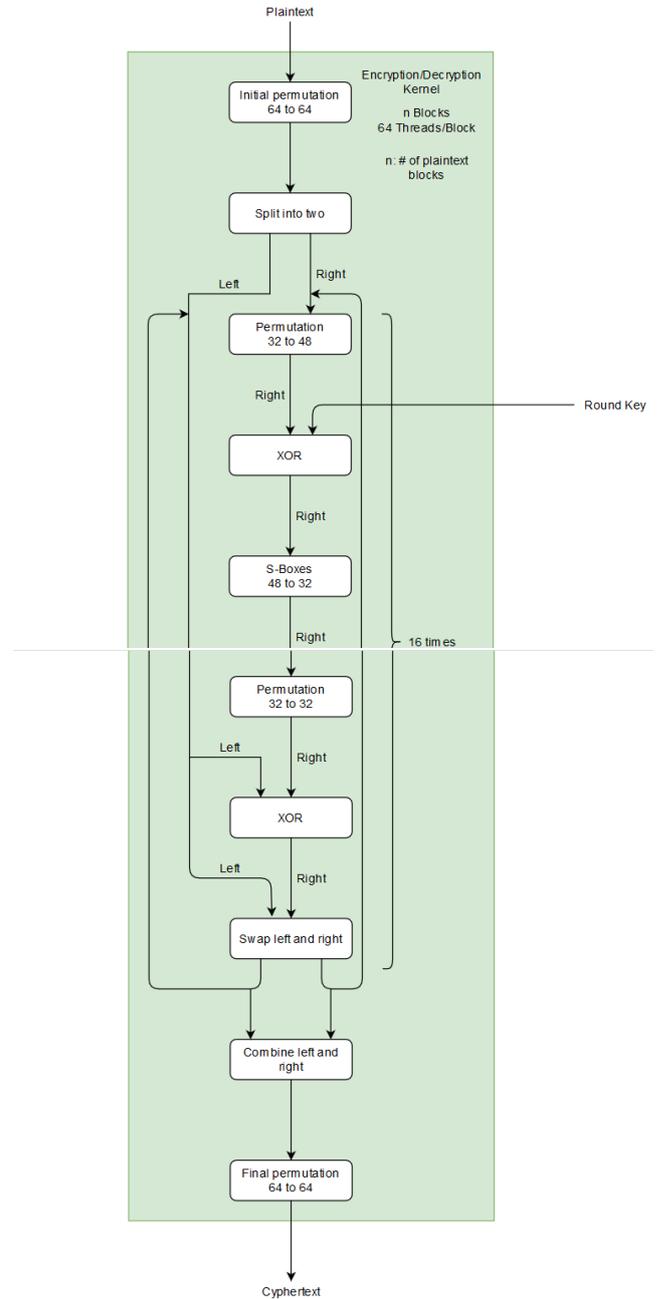

Figure 3: Encryption/decryption kernel flowchart.

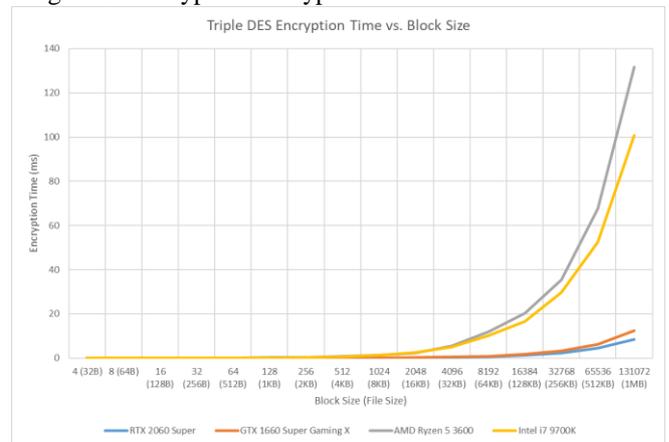

Figure 4: 3DES encryption time vs. block size graph for different processors.

*Table I: Encryption time (ms) for different processors*

| Block Size (File Size) | GPU-1 | GPU-2 | CPU-1 | CPU-2 |
|---|---|---|---|---|
| 4 (32B) | 0.032 | 0.029 | 0.012 | 0.017 |
| 8 (64B) | 0.033 | 0.030 | 0.014 | 0.016 |
| 16 (128B) | 0.033 | 0.030 | 0.034 | 0.028 |
| 32 (256B) | 0.032 | 0.030 | 0.055 | 0.050 |
| 64 (512B) | 0.033 | 0.031 | 0.107 | 0.096 |
| 128 (1KB) | 0.032 | 0.031 | 0.209 | 0.184 |
| 256 (2KB) | 0.035 | 0.037 | 0.406 | 0.367 |
| 512 (4KB) | 0.050 | 0.064 | 0.718 | 0.653 |
| 1024 (8KB) | 0.089 | 0.120 | 1.387 | 1.237 |
| 2048 (16KB) | 0.161 | 0.225 | 2.401 | 2.463 |
| 4096 (32KB) | 0.305 | 0.438 | 5.498 | 4.977 |
| 8192 (64KB) | 0.592 | 0.862 | 11.986 | 10.144 |
| 16384 (128KB) | 1.184 | 1.725 | 20.354 | 16.701 |
| 32768 (256KB) | 2.324 | 3.396 | 35.319 | 29.843 |
| 65536 (512KB) | 4.511 | 6.261 | 67.626 | 52.587 |
| 131072 (1MB) | 8.399 | 12.441 | 131.577 | 100.715 |

*Table II: GPU vs CPU Speed-ups*

| Block Size (File Size) | GPU-1 vs CPU-1 | GPU-2 vs CPU-1 | GPU-1 vs CPU-2 | GPU-2 vs CPU-2 |
|---|---|---|---|---|
| 4 (32B) | 0.38 | 0.41 | 0.53 | 0.58 |
| 8 (64B) | 0.42 | 0.47 | 0.49 | 0.53 |
| 16 (128B) | 1.03 | 1.13 | 0.85 | 0.93 |
| 32 (256B) | 1.72 | 1.83 | 1.56 | 1.67 |
| 64 (512B) | 3.24 | 3.45 | 2.91 | 3.10 |
| 128 (1KB) | 6.53 | 6.74 | 5.75 | 5.94 |
| 256 (2KB) | 11.60 | 10.97 | 10.48 | 9.92 |
| 512 (4KB) | 14.36 | 11.22 | 13.06 | 10.20 |
| 1024 (8KB) | 15.59 | 11.56 | 13.89 | 10.30 |
| 2048 (16KB) | 14.91 | 10.67 | 15.29 | 10.94 |
| 4096 (32KB) | 18.03 | 12.55 | 16.31 | 11.36 |
| 8192 (64KB) | 20.25 | 13.91 | 17.13 | 11.76 |
| 16384 (128KB) | 17.20 | 11.80 | 14.10 | 9.68 |
| 32768 (256KB) | 15.20 | 10.40 | 12.84 | 8.79 |
| 65536 (512KB) | 15.00 | 10.80 | 11.65 | 8.40 |
| 131072 (1MB) | 15.67 | 10.58 | 12.00 | 8.10 |

*Table III: Average speed-ups of the GPUs with respect to CPUs*

|  | CPU-1 | CPU-2 |
|---|---|---|
| GPU-1 | 10.70 | 9.31 |
| GPU-2 | 8.03 | 7.01 |

APPENDIX A: CONSTANT ARRAYS

```
//Permuted choice table
BYTE pc_1[56] =
{
  57,49,41,33,25,17,9,
  1,58,50,42,34,26,18,
  10,2,59,51,43,35,27,
  19,11,3,60,52,44,36,
  63,55,47,39,31,23,15,
  7,62,54,46,38,30,22,
  14,6,61,53,45,37,29,
  21,13,5,28,20,12,4
};

int shift_keys[16] =
{
  1, 1, 2, 2,
  2, 2, 2, 2,
  1, 2, 2, 2,
  2, 2, 2, 1
};

//Key-Compression Table
BYTE pc_2[48] =
{
  14,17,11,24,1,5,
  3,28,15,6,21,10,
  23,19,12,4,26,8,
  16,7,27,20,13,2,
  41,52,31,37,47,55,
  30,40,51,45,33,48,
  44,49,39,56,34,53,
  46,42,50,36,29,32
};

//Initial Permutation
BYTE initial_perm[64] =
{
  58,50,42,34,26,18,10,2,
  60,52,44,36,28,20,12,4,
  62,54,46,38,30,22,14,6,
  64,56,48,40,32,24,16,8,
  57,49,41,33,25,17,9,1,
  59,51,43,35,27,19,11,3,
  61,53,45,37,29,21,13,5,
  63,55,47,39,31,23,15,7
};

//Expansion D-box Table
BYTE exp_d[48] =
{
  32,1,2,3,4,5,4,5,
  6,7,8,9,8,9,10,11,
  12,13,12,13,14,15,16,17,
  16,17,18,19,20,21,20,21,
  22,23,24,25,24,25,26,27,
  28,29,28,29,30,31,32,1
};

//S-box Table, total 8 s-boxes
BYTE s[8][4][16] =
{ {
  14,4,13,1,2,15,11,8,3,10,6,12,5,9,0,7,  //0
  0,15,7,4,14,2,13,1,10,6,12,11,9,5,3,8,
  4,1,14,8,13,6,2,11,15,12,9,7,3,10,5,0,
  15,12,8,2,4,9,1,7,5,11,3,14,10,0,6,13
},
{
  15,1,8,14,6,11,3,4,9,7,2,13,12,0,5,10,  //1
  3,13,4,7,15,2,8,14,12,0,1,10,6,9,11,5,
  0,14,7,11,10,4,13,1,5,8,12,6,9,3,2,15,
  13,8,10,1,3,15,4,2,11,6,7,12,0,5,14,9
},
{
  10,0,9,14,6,3,15,5,1,13,12,7,11,4,2,8,  //2
  13,7,0,9,3,4,6,10,2,8,5,14,12,11,15,1,
  13,6,4,9,8,15,3,0,11,1,2,12,5,10,14,7,
  1,10,13,0,6,9,8,7,4,15,14,3,11,5,2,12
},
{
  7,13,14,3,0,6,9,10,1,2,8,5,11,12,4,15,  //3
  13,8,11,5,6,15,0,3,4,7,2,12,1,10,14,9,
  10,6,9,0,12,11,7,13,15,1,3,14,5,2,8,4,
  3,15,0,6,10,1,13,8,9,4,5,11,12,7,2,14
},
{
  2,12,4,1,7,10,11,6,8,5,3,15,13,0,14,9,  //4
  14,11,2,12,4,7,13,1,5,0,15,10,3,9,8,6,
  4,2,1,11,10,13,7,8,15,9,12,5,6,3,0,14,
  11,8,12,7,1,14,2,13,6,15,0,9,10,4,5,3
```

```
        },
        {
            12,1,10,15,9,2,6,8,0,13,3,4,14,7,5,11,   //5
            10,15,4,2,7,12,9,5,6,1,13,14,0,11,3,8,
            9,14,15,5,2,8,12,3,7,0,4,10,1,13,11,6,
            4,3,2,12,9,5,15,10,11,14,1,7,6,0,8,13
        },
        {
            4,11,2,14,15,0,8,13,3,12,9,7,5,10,6,1,   //6
            13,0,11,7,4,9,1,10,14,3,5,12,2,15,8,6,
            1,4,11,13,12,3,7,14,10,15,6,8,0,5,9,2,
            6,11,13,8,1,4,10,7,9,5,0,15,14,2,3,12
        },
        {
            13,2,8,4,6,15,11,1,10,9,3,14,5,0,12,7,   //7
            1,15,13,8,10,3,7,4,12,5,6,11,0,14,9,2,
            7,11,4,1,9,12,14,2,0,6,10,13,15,3,5,8,
            2,1,14,7,4,10,8,13,15,12,9,0,3,5,6,11
        } };

    //Straight Permutation Table
    BYTE per[32] =
    {
        16,7,20,21,
        29,12,28,17,
        1,15,23,26,
        5,18,31,10,
        2,8,24,14,
        32,27,3,9,
        19,13,30,6,
        22,11,4,25
    };

    //Final Permutation Table
    BYTE final_perm[64] =
    {
        40,8,48,16,56,24,64,32,
        39,7,47,15,55,23,63,31,
        38,6,46,14,54,22,62,30,
        37,5,45,13,53,21,61,29,
        36,4,44,12,52,20,60,28,
        35,3,43,11,51,19,59,27,
        34,2,42,10,50,18,58,26,
        33,1,41,9,49,17,57,25
    };
```